# Study of the 27-day variations in GCR fluxes during 2007-2008 based on PAMELA and ARINA observations

R. Modzelewska[1], G.A. Bazilevskaya[2], M. Boezio[3,4], S.V. Koldashov[5], M.B. Krainev[2], N. Marcelli[6,7], A.G. Mayorov[5], M.A. Mayorova[5], R. Munini[3,4], , I.K. Troitskaya[5], R.F. Yulbarisov[5], X. Luo[8], M.S. Potgieter[9], O.P.M. Aslam[10]

**Abstract**

Using measurements from the *PAMELA* and *ARINA* spectrometers onboard the RESURS DK-1 satellite, we have examined the 27-day intensity variations in galactic cosmic ray (GCR) proton fluxes in 2007-2008. The *PAMELA* and *ARINA* data allow for the first time a study of time profiles and the rigidity dependence of the 27-day variations observed directly in space in a wide rigidity range from ~300 MV to several GV. We find that the rigidity dependence of the amplitude of the 27-day GCR variations cannot be described by the same power-law at both low and high energies. A flat interval occurs at rigidity $R = <0.6\text{-}1.0>$ GV with a power-law index $\gamma = -0.13\pm0.44$ for *PAMELA*, whereas for $R \geq 1$ GV the power-law dependence is evident with index $\gamma = -0.51\pm0.11$. We describe the rigidity dependence of the 27-day GCR variations for *PAMELA* and *ARINA* data in the framework of the modulation potential concept using the force-field approximation for GCR transport. For a physical interpretation, we have considered the relationship between the 27-day GCR variations and solar wind plasma and other heliospheric parameters. Moreover, we have discussed possible implications of MHD modeling of the solar wind plasma together with a stochastic GCR transport model concerning the effects of corotating interaction regions.

[1] Corresponding author renatam@uph.edu.pl
[1] Institute of Mathematics, Siedlce University, Konarski Str. 2 08110 Siedlce, Poland
[2] Lebedev Physical Institute, RAS, RU-119991 Moscow, Russia
[3] INFN, Sezione di Trieste, I-34149 Trieste, Italy
[4] IFPU, I-34014 Trieste, Italy
[5] National Research Nuclear University MEPhI, RU-115409 Moscow, Russia
[6] University of Rome "Tor Vergata", Department of Physics, I-00133 Rome, Italy
[7] INFN, Sezione di Rome "Tor Vergata", I-00133 Rome, Italy
[8] Shandong Institute of Advanced Technology, 250100 Jinan, China
[9] Retired; FS4, Potchefstroom, South Africa
[10] North-West University, Centre for Space Research, 2520 Potchefstroom, South Africa

## 1. Introduction

The recurrent variations of the galactic cosmic ray (GCR) intensity and anisotropy which are due to the passage through the point of measurement of the solar wind (SW) and heliospheric magnetic field (HMF) structures rotating with the Sun, have been studied for more than 60 years (Simpson 1998). According to the modern concept, the source of these structures is the longitudinal gradient of the SW velocity in the heliosphere near the Sun, connected in turn with the geometry of the flux-tube from the coronal holes into the heliosphere (Wang & Sheeley 1990; Arge & Pizzo 2000). As a result, in the inner heliosphere ($r < 1$ AU) the stream interaction regions (SIRs) are formed between the low-velocity stream and the overtaking fast-velocity stream originated from coronal holes on the Sun. The interaction region due to the rotation of the Sun, is twisted approximately into a Parker spiral. Due to the long lived coronal holes rotating with the Sun, this structure is seen by an observer as a periodic corotating interaction region (CIR) (Richardson 2018). At larger distances ($r \approx 3-6$ AU) this structure of the compressed HMF is expanding with two shock-waves around the CIR. In the case of a stable position and power of the SIR's source near the Sun for 5-10 solar rotations (the synodic solar period $P_\odot \approx 27$ days), a CIR spreads over several AU, and a 27-day variation arises in the GCR persisting over several months. The shocks connected with CIR may accelerate particles up to 20 MeV/nucleon (for a review see Richardson 2004, 2018). CIRs are especially prominent during the declining phase of the solar cycle and occur usually at low latitudes, where the HMF has a well-established sector structure and coronal holes spread to low helio-latitudes. Such a situation is characteristic of periods near solar cycle minima. Consequently, the 27-day GCR variations are generally more evident and typical with longer duration during the minimum and near minimum epochs of solar activity. These 27-day variations of GCRs are observed not only by ground based neutron monitors (NMs) e.g., Gil & Mursula (2017), but also near Earth by space missions e.g. *IMP8* (Richardson et al. 1996), *ACE* (Leske et al. 2011, 2013, 2019) and others, in the inner heliosphere also at high helio-latitudes on *Ulysses* (McKibben et al. 1995; Heber et al. 1999; see review by Heber & Potgieter 2006), and even on the *Voyager* spacecraft (Decker et al. 1999) in the outer heliosphere, confirming them to be extensive in scope.

Despite the long history of observations, some characteristic features of the 27-day variations are still not known well enough (Simpson 1998; Richardson 2018), for example, their HMF polarity dependence and, in particular, their dependence on the particle's rigidity over a wide rigidity (energy) range.

The polarity dependence of the 27-day amplitude of GCRs, A27, (Richardson et al. 1999, Alania et al. 2008), was lately experimentally confirmed by Gil & Mursula (2017) using the Apatity and Oulu NMs. They explained it by a combination of drift effects and SW convection, showing that in the period of negative polarity (A<0) amplitudes are smaller because the heliospheric current sheet (HCS) plays a dominant role with convection effects small. On the other hand, during periods of positive polarity (A>0) amplitudes are larger due to GCRs drifting from polar to equatorial regions over a wider range of helio-latitudes and meeting at higher helio-latitudes faster SW. These HMF polarity dependent trends in these amplitudes are however in contrast to what is predicted for Forbush decreases by Luo et al. (2017). Additionally, Gil & Mursula (2017) reported a diminishing trend in the amplitudes of these recurrent variations connected to the Sun's rotation during consecutive solar minima, which can be associated with the weakening in the solar polar magnetic fields during the last four solar cycles. However, Modzelewska & Alania (2012) and Gil et al. (2012) showed that based on Kiel and Moscow NM observations the amplitude of the 27-day variation in 2007-2008 was comparable to that of 1995-1997. Additionally, Modzelewska & Alania (2012) attributed the polarity dependence of the 27-day variations in the mid-80s and 90s to the larger amplitudes of the azimuthal changes of the SW velocity in the A>0 polarity for mid-90's. Observations of the 27-day GCR variations in the 2007-2008 period show the associated stable recurrent variations with the period of solar rotation (~27 days) also in solar wind velocity, with amplitudes comparable or even higher in comparison to the previous period of positive polarity (1995-1997). Evidently, this topic needs further study. So, we discuss the behavior of the 27-day GCR variations in 2007-2008 and compare it with 1996-1997 using only the periods of enhanced periodic variability of the GCR intensity, not the average amplitude during the whole consecutive minima.

The rigidity dependence of A27, was studied in a sequence of publications by Gil & Alania (2010, 2011, 2013, 2016), showing that the spectrum was a power-law. Gil & Alania (2016) demonstrated that the power-law rigidity spectrum of the recurrent variations of the GCR intensity is harder during maximum epochs, and softer during the minimum epochs of solar activity. **It was suggested by Gil & Alania (2010) that this phenomenon could be related to changes in the extent of the heliospheric regions over which these recurrent variations of GCR intensity occur during different epochs of solar activity. The mentioned authors considered the region of these recurrences to be smaller during minimum epochs than during maximum epochs.** However, these studies concerned only the energy range covered by observations with NMs, that is, for rigidities $R > 10$ GV.

The study of GCR recurrent variations during 1992-1993 onboard *Ulysses* revealed a maximum in the rigidity dependence of A27 around 1 GV (McKibben et al. 1995). The recurrent GCR variations were seen from the equatorial to high helio-latitudes; and a linear relationship between the GCR latitudinal gradient and A27 was reported by Zhang (1997). This implies the existence of a modulation mechanism controlling both the global latitudinal distribution and the short-term temporal variation of GCR fluxes. Paizis et al. (1997) described the rigidity dependence of A27 observed by *Ulysses* in 1992-1993 in the framework of the modulation potential formalism as introduced by Gleeson & Axford (1968). The concept of Zhang (1997) was not confirmed during the third high-latitude scan of *Ulysses* in 2005-2006 which also took place near solar minimum but with the opposite HMF polarity (Dunzlaff et al. 2008). Contrary to the first *Ulysses* high-latitude scan (1992-1994), no latitude dependence of A27 was seen up to 40 degrees and the periodic GCR modulation was absent at higher latitudes although the recurrent SW patterns persisted. Kota & Jokipii (1991) first proposed a model for these recurrent variations with drifts and CIRs, predicting larger depressions for A<0 polarity cycles, which turned out to be inconsistent with observations of Richardson et al. (1999), probably because the predicted effects of drifts were vastly overestimated. Next, Kota & Jokipii (2001a), using a 3D GCR transport model with a southward shift of the HCS and CIRs, could reproduce the polarity dependence of A27. The polarity dependence of this amplitude was explained also using a Fisk-type HMF (Burger et al. 2008), but controversy has persisted with opposite views about the existence and actual effects of such a Fisk field (Fisk 1996) near solar minimum (see e.g. Roberts et al. 2007; Sternal et al. 2011). Dunzlaff et al. (2008) attributed the different characteristic features of the 27-day GCR variations during the two high-latitude *Ulysses* scans to the difference in the coronal hole structures between cycles 22 (1986 – 1996) and 23 (1996 – 2008); an extended, stable coronal hole structure was present during cycle 22, but not in cycle 23. The polar coronal hole disappeared during the declining phase of cycle 23 (Kirk et al. 2009) and a part of the coronal hole structure existed in the equatorial region (Abramenko et al. 2010).

It has become clear that the prominent 27-day GCR variations near the minimum of cycle 23 actually developed in 2007-2008, after the period of 2005-2006 analyzed by Dunzlaff et al. (2008). Owing to the wide energy range of GCRs provided by the *PAMELA* mission (Picozza et al. 2007; Adriani et al. 2014; Martucci et al. 2018) we now have an opportunity to retrieve the rigidity spectrum of the amplitude of the recurrent GCR variations in 2007-2008 and to compare it with the result obtained by *Ulysses* in 1992-1993.

The clear existence of the stable 27-day GCR variations in 2007-2008 has inspired researchers in a discussion about modulation processes governing these phenomena from a theoretical point of view. In a sequence of publications Alania et al. (2010, 2011) and Modzelewska & Alania (2013) studied theoretically these 27-day variations and successfully reproduced them as observed by NMs during 2007-2008. Their numerical model was based on Parker transport equation (Parker 1965) that incorporates the observed recurrent changes of the SW speed and corresponding consistent divergence-free HMF. In alternative approaches, Guo & Florinski (2014, 2016), Wiengarten et al. (2014) and Kopp et al. (2017) used MHD modeling of the SW and HMF serving as input to a GCR transport code employing a stochastic differential equation approach (Zhang 1999). Guo & Florinski (2014, 2016) found that GCR variations were dominated by the HCS when its tilt angle became small and the depressions in the GCR intensity were directly caused by the longitudinal and radial changes in diffusion coefficients passing from the slow to the fast solar wind, and vice versa, associated with the passages of stream interfaces. They concluded that the recurrent GCR variations in 2007-2008 were relatively independent of the HMF magnitude and were unrelated to the sector boundary/current sheet crossings. In spite of undoubted progress, this model has some inconsistency when compared with observations, e.g. showing that the GCR variations are often deepest near the stream leading edge, then recover during stream passage (Richardson 2018).

Although a generic connection between CIRs and the 27-day variation in GCR is obvious, the mechanism of the 27-day GCR variation is not quite clear because of the multiplicity and complexity of processes involved. The timings of the CIR structures and the GCR flux modulation are closely connected. However, no particular modulation parameter was found to be solely responsible or being a decisive factor for GCR modulation (Kumar & Badruddin 2014). It is fair to say that the 27-day variation is still not well reproduced by theoretical/numerical models (e.g., Guo & Florinski 2014, 2016). Actually, the relative roles of various modulation mechanisms are now the most interesting aspect in the understanding of 27-day GCR variations.

Our paper is devoted to the well-known episode of the 27-day GCR variations in 2007-2008, near the minimum of solar cycle 23, the period exclusively favorable for the development of the pronounced and long-lived recurrent GCR variations. This episode was extensively studied based on observations with NMs (e.g., Modzelewska & Alania 2013; Gil & Alania 2016; Gil & Mursula 2018) and space probes (e.g. Gieseler et al. 2009; Leske et al. 2011; Krainev et al. 2018). All authors emphasized the very stable period of ~ 27 days and a strong negative correlation of the 27-day waves in GCRs and SW velocity. Correlations with the HMF strength and its components were less prominent, however. Also, for 2007-2008, Leske et al. (2011) reported particle enhancements, accelerated by CIRs, observed at 1 AU during the long and deep solar minimum of 2007-2009, practically free of solar energetic particle (SEP) contamination.

Gil and Mursula (2018) and Krainev et al. (2018) compared the two episodes, 2007-2008 and 2014-2015, of the 27-day variations in solar cycle 24. Gil and Mursula (2018) suggested that in both cases the source of periodic variations was a coronal hole, but the coronal holes in the two intervals were located quite differently: a strongly north-south asymmetric polar coronal hole existed in 2014-2015 and a trans-equatorial coronal hole governed the periodic variations during 2007-2008. Additionally, these authors mentioned that the differences in recurrent modulation effects take place in opposite solar magnetic polarity epochs; that is, A < 0 for 2007-2008 but A > 0 for 2014-2015. Krainev et al. (2018) demonstrated that the situation in 2007-2008, unlike that in 2014-2015, was rather stable and consistent with the classic picture of a stable CIR and corresponding GCR variation in it, as is described by Richardson (2004, 2018).

Recently, Ghanbari et al. (2019) studied the turbulence properties around CIRs during the two recent solar minima, 2007-2008 and 2017-2018. They found that, similarly for both periods, the maximum of the total turbulent energy occurs half a day after the stream interface (SI), separating the fast and slow SW, with 2 times greater energy for the fast wind than for slow wind. They noted the higher levels of turbulence in the fast wind during 2007-2008. Using a superposed method they found a significant correlation between proton count rate for kinetic energy (KE) >120 MeV from ACE (CRIS) and the perpendicular diffusion coefficient for GCRs, suggesting that the perpendicular transport is the main source of GCR modulation around CIRs.

Here, we present the results of the 27-day variation in GCR protons with rigidities from ~0.3 GV to ~10 GV, (KE from ~0.05 to ~10 GeV) as observed by the space-borne instruments *PAMELA* and *ARINA* in 2007-2008. *PAMELA* observations fill the largely unexplored energy gap between the GCR particles detected in space (below a few hundred MeV) and particles detected on the Earth (KE >10 GeV). *PAMELA* and *ARINA* data allow for the first time to study the rigidity dependence of the 27-day variation of GCRs observed directly in space over a wide rigidity range so that it is possible to investigate the time and rigidity profiles of these GCR intensity variations.

We study the rigidity dependence of A27 and the relation of periodic GCR variations with heliospheric parameters. We discuss the results in the framework of known GCR modulation mechanisms.

The structure of the paper is as follows: In Section 2 the *PAMELA* and *ARINA* experiments are briefly described. The daily GCR intensities and their time behavior are discussed and the period of intense 27-day GCR variations is isolated and shown in Section 3. Section 4 is devoted to the overall description of the situation about the 27-day variations in terms of heliospheric characteristics and the GCR intensity in the mentioned 2007-2008 period. In Section 5, we present the method of processing and then results on the rigidity dependence of the amplitude of the 27-day variations in the proton intensity. This is followed by a discussion in Section 6 and conclusions in Section 7.

## 2. *PAMELA* and *ARINA* experiments

The spectrometers *PAMELA* (Adriani et al. 2014) and *ARINA* (Bakaldin et al. 2007) situated on the same spacecraft Resurs DK1, had been operational for almost 10 years since June 2006. In 2007-2008, the satellite orbit was elliptical (altitude varying between 355 and 584 km) with inclination of about 70° and a period of about 94 minutes. The instrument allowed the measurement of protons, electrons, their antiparticles, and light nuclei in the KE interval from several tens of MeV up to several hundreds of GeV. The instrument consisted of a magnetic spectrometer with a silicon tracking system, a time-of-flight system shielded by an anticoincidence system, an electromagnetic calorimeter and a neutron detector. The data treatment is described in detail by Munini et al. (2017).

The *ARINA* telescope was a multilayer scintillation detector consisting of 10 plates arranged as a truncated pyramid. Particles were identified by the energy loss in each detector and the path until stopping measured in the number of plates (*dE/dX* vs *E* method). The instrument detected electrons with energies of 3-30 MeV and protons with energies of 30-110 MeV. The energy resolution of the ARINA spectrometer was 10-15%. The aperture of the device was ~ 10 $cm^2$ sr. For the extraction of the galactic component, events with energy higher than the geomagnetic cutoff were selected at each registration point, on L-shells no less than 8.

Here we concentrate on the investigation of the periodic variation of proton fluxes measured by *PAMELA* and *ARINA* in 2007-2008 taking place during the prolonged solar minimum between solar cycles 23 and 24. We compare the *PAMELA* and *ARINA* observations with NMs data.

We use the daily proton fluxes:
- PAMELA; 15 rigidity bins:
  0.40-0.47 GV; 0.47-0.55 GV; 0.55-0.64 GV; 0.64-0.75 GV; 0.75-0.88 GV; 0.88-1.03 GV; 1.03-1.20 GV; 1.20-1.54 GV; 1.54-1.99 GV; 1.99-2.56 GV; 2.56-3.29 GV; 3.29-4.23 GV; 4.23-5.44; 5.44-7.00 GV; 7.00-10.37 GV;
- ARINA; 7 rigidity bins:
  0.29-0.32 GV; 0.31-0.34 GV; 0.33-0.37 GV; 0.36-0.41 GV; 0.39-0.42 GV; 0.41-0.44 GV; 0.44-0.46 GV;

Daily proton fluxes for selected rigidities from *PAMELA* and *ARINA* and the Oulu NM daily count rate in 2007-2008 are presented in Fig. 1, respectively. Data from 12-30 September 2008 are missing from observations.

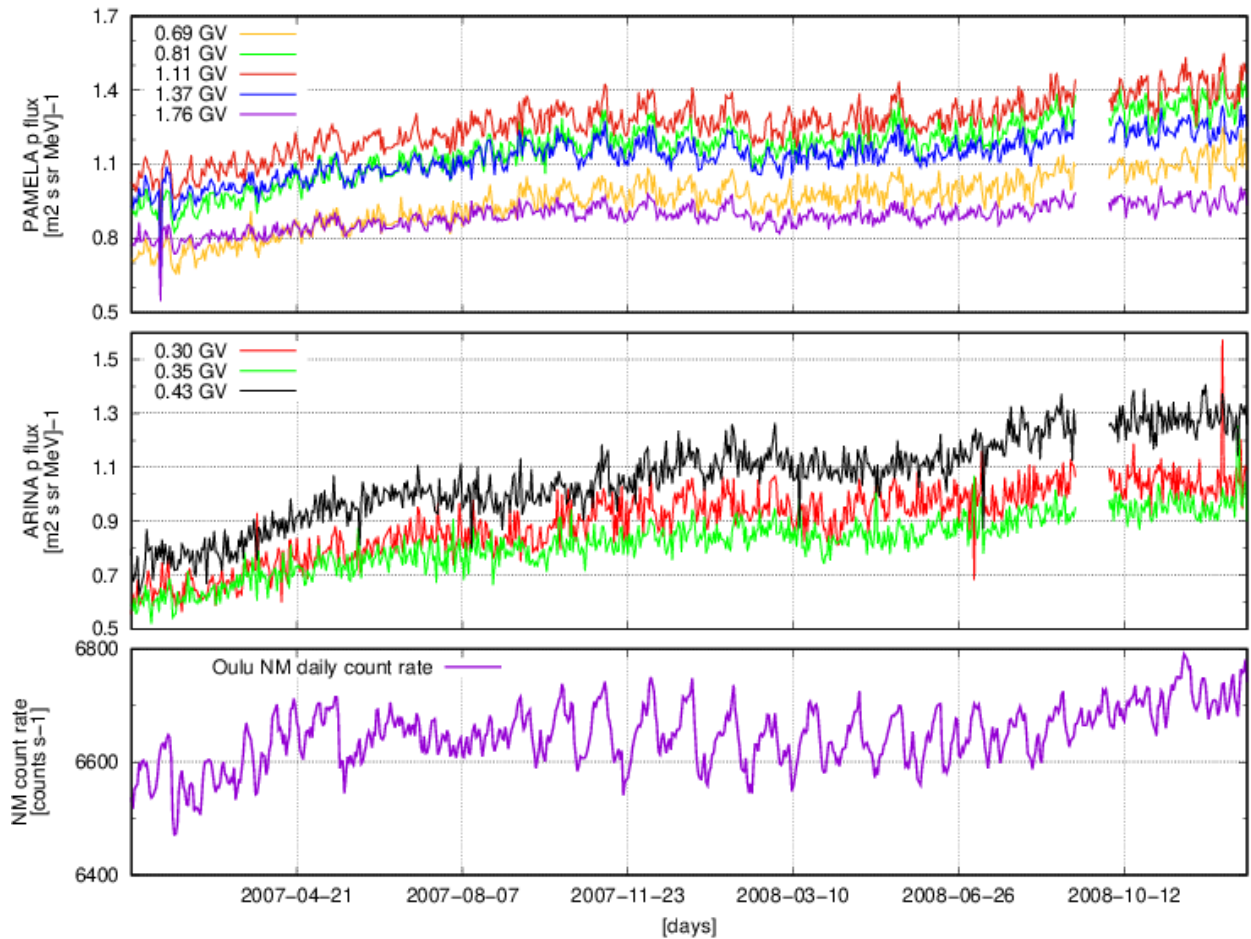

**Figure 1** Daily proton fluxes for a selected range of rigidities for *PAMELA* (upper panel), and for *ARINA* (middle panel), and the Oulu NM (cutoff rigidity 0.8 GV) daily count rate (lower panel),

in 2007-2008.

## 3. Data treatment

To study the 27-day GCR variation, daily proton fluxes of *PAMELA* and *ARINA* were detrended as $\frac{(x-\tilde{x}_{29d})}{\tilde{x}_{29d}} \cdot 100\%$, where $\tilde{x}_{29d}$ is the running 29 day average. Relative proton fluxes, 5 days running average, are presented for *PAMELA* (top panel in Fig. 2a), *ARINA* (top panel in Fig.2b) and the Oulu NM (top panel in Fig. 2c), respectively. **The colors are the same as in Figure 1.**

To study dynamics of the temporal changes of the periodicity connected with the Sun's rotation the wavelet time-frequency spectrum technique was used. In our calculation we adopted the Morlet wavelet mother function (Torrence & Compo 1998). As examples we present the results of wavelet analysis of the proton flux by *PAMELA* for rigidity 1.8 GV (bottom panel in Fig. 2a), *ARINA* for rigidity 0.43 GV (bottom panel in Fig. 2b) and the Oulu NM (bottom panel in Fig. 2c).

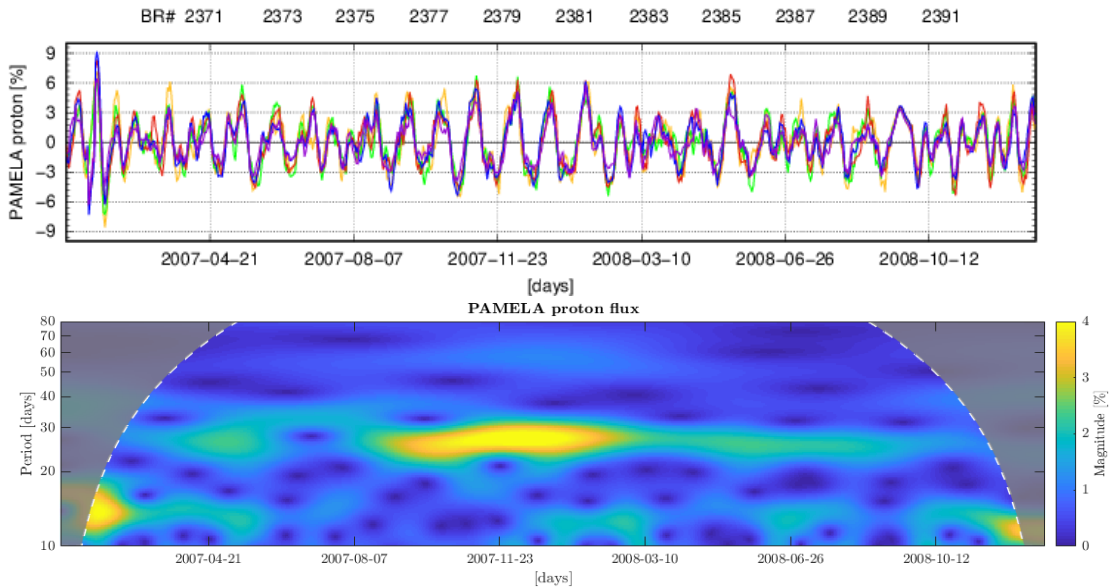

**Figure 2a** (top) Temporal evaluation of the relative proton fluxes, 5-day running averages from Fig. 1 measured by *PAMELA,* **colors are the same as in Figure 1**; (bottom) wavelet analysis **of daily *PAMELA* proton flux for rigidity 1.8 GV** for 2007-2008; dashed line corresponds to the 95% confidence level.

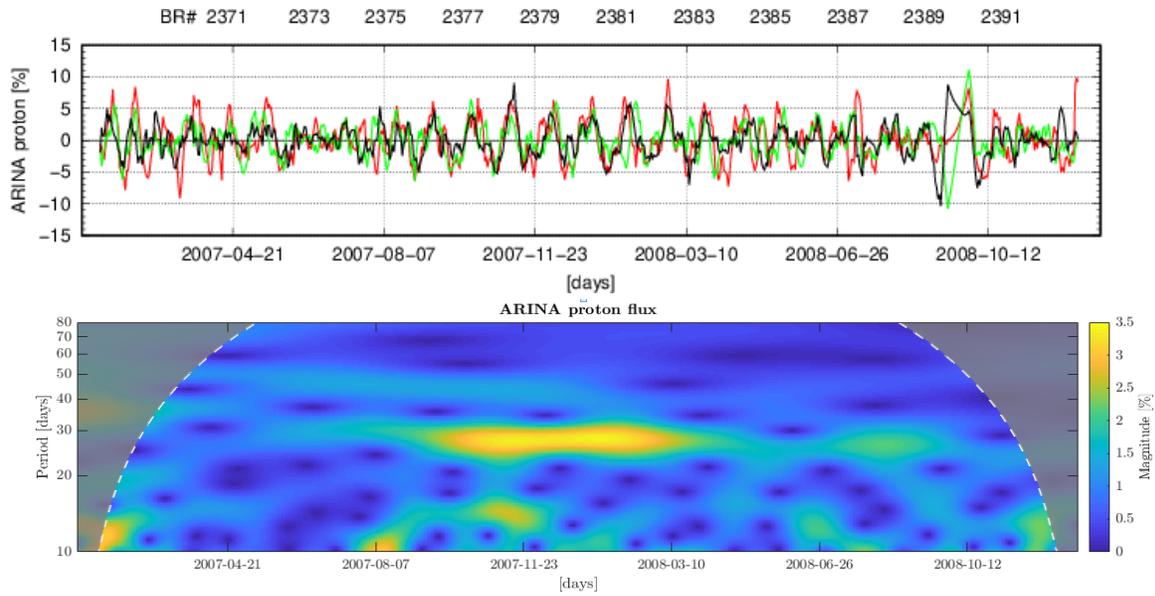

**Figure 2b** Similar to Fig. 2a but for *ARINA* data **for rigidity 0.43 GV**.

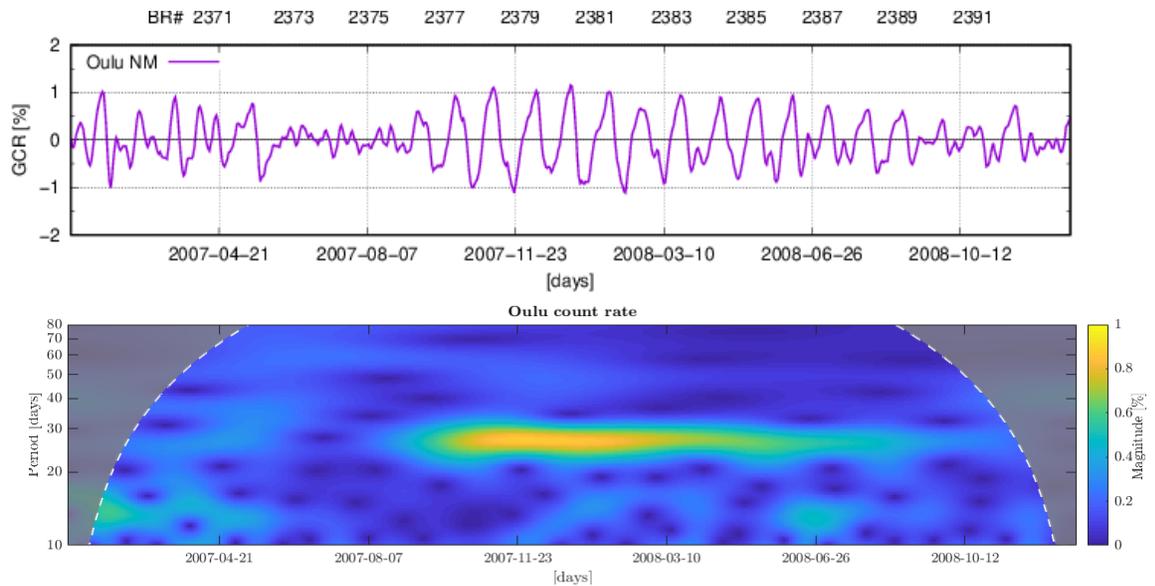

**Figure 2c** Similar to Fig. 2a but for the Oulu NM.

Using the method adopted by Modzelewska & Alania (2013) we calculated the power of the 27-day variations for *PAMELA* and *ARINA* proton fluxes and the Oulu NM presented in Fig. 3.

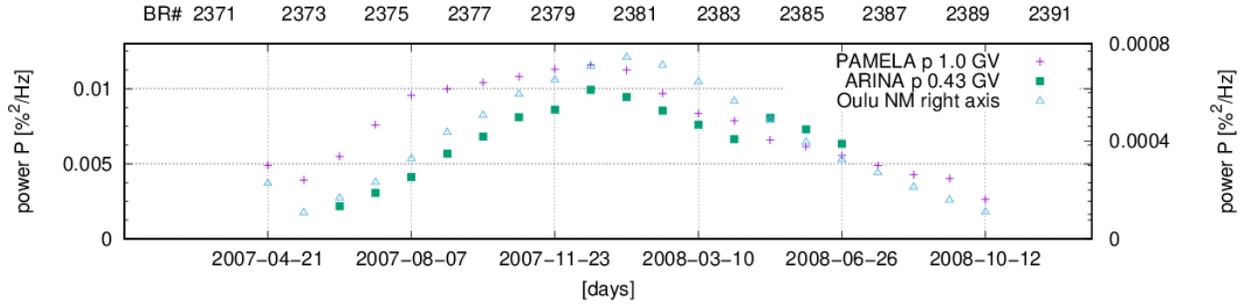

**Figure 3** Power *P* of the recognized periodicity of ~27-days for *PAMELA*, *ARINA* and the Oulu NM for 2007-2008.

The results presented confirm the high power and large amplitude of the 27-day GCR variations in the period of 30 September 2007 – 11 February 2008 corresponding to Bartels' rotations (BRs) 2377-2381. We choose this time interval for further analysis.

Since the main source of the 27-day GCR variations is a corotating interaction region (CIR), we also analyze the relevant SW parameters: daily *Br* component of the HMF and solar wind velocity *V* [OMNI]. In Figure 4 we present wavelet analysis and power of the 27-day variation of the daily *Br* component and *V* for 2007-2008. The dominant amplitude of the 27-day variation for this *Br* component and *V* coincides with the *PAMELA* and *ARINA* data, but stable periodic variations in *V* are clearly starting earlier and for *Br* component lasting longer than in the GCRs.

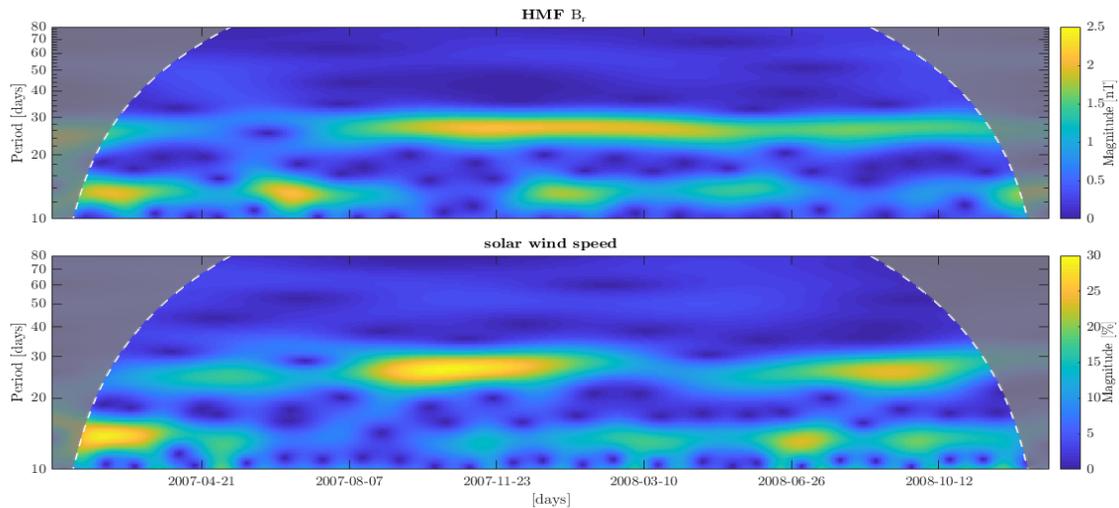

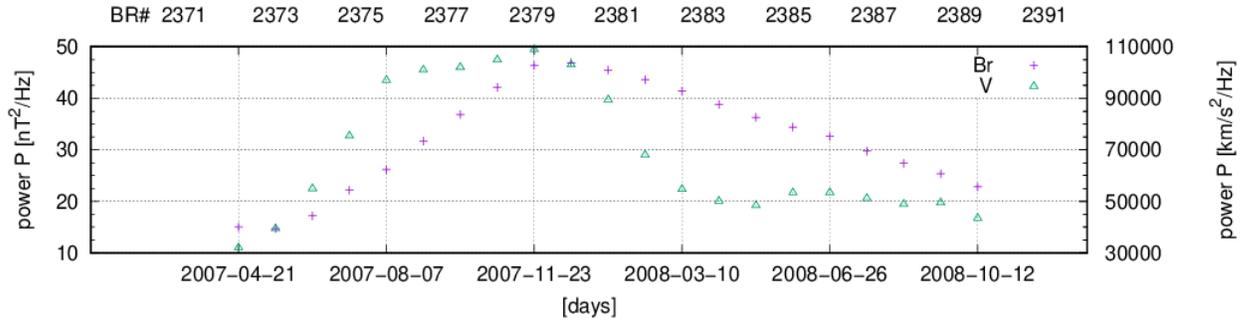

**Figure 4** Results of wavelet analysis (top) and the power of the 27-day wave (bottom) of the daily *Br* component of the HMF and SW velocity *V* for 2007-2008.

4. **Heliospheric and GCR characteristic features in the period of intense 27-day variations during 2007-2008**

Starting from August 2007 the SW velocity demonstrated the pronounce 27-day variation with which the 27-day GCR variations obviously were connected. Fig. 5 gives a comparison of the 27-day variations in GCR, in the SW, and the HMF parameters for the Bartels rotations 2377-2381 [OMNI]. All data are in de-trended form with an exception for the HMF *Br* component which is given as itself to emphasize the times of the HCS crossing. There are two distinct peaks in the SW velocity divided by two valleys of quite different duration: a short valley at the end of each BR, correlating with the fast crossing of the low-velocity layer surrounding the HCS, and the long valley in the velocity in the middle part of each BR associated with the slow crossing of the HCS layer only slightly tilted to the Earth's trajectory. The heliospheric conditions were dominated by SIRs (shadowed bands in Fig. 5, according to Jian et al. (2006a, 2011) which bear main characteristic SIR features, namely, the positive SW speed derivative and relative peaks in the SW density, temperature, and HMF strength. During this period, there were only two weak interplanetary coronal mass ejecta (ICMEs) (Jian et al. 2006b, 2011) (red bands in Fig. 5) on 19.11.2007.
[http://www.srl.caltech.edu/ACE/ASC/DATA/level3/icmetable2.htm] and 25.12. 2007
[https://wind.nasa.gov/2007.php ].
So, the time behavior of the SW, HMF and GCR intensity near the Earth in the selected period of the intense 27-day variations can be considered as the manifestation of the steady but longitudinally dependent variations in the coordinate system rotating with the Sun. On the other hand, the GCR intensity for both lower and higher rigidities demonstrates a smoother 27-day wave. There is one broad minimum in the GCR intensity formed around two close velocity peaks (**arrows in Fig. 5 indicate the double SW velocity peaks**) and a broad maximum around the long valley in the velocity. **Fig. 5 shows that there is no one-to-one correspondence between the variations in the GCR intensity and in the heliospheric characteristics measured near the Earth. Particularly double peaks in SW are not seen in GCR intensity as the same percentage rate (it will be discussed in detail later in Fig. 6) and there is no correlation between deviations from the averages of HMF magnitude and relative GCR** intensities, correlation coefficient $\rho$ between $\varepsilon J_{low}$ and $\varepsilon B$ is $\rho=-0.12\pm0.01$, $\varepsilon J_{high}$ and $\varepsilon B$ $\rho=-0.22\pm0.01$, $\varepsilon J_{low}$ and $Br$ $\rho=-0.14\pm0.01$, $\varepsilon J_{high}$ and $Br$ $\rho=-0.16\pm0.01$, while $\varepsilon J_{low}$ and $\varepsilon V$ $\rho=-0.70\pm0.01$, $\varepsilon J_{high}$ and $\varepsilon V$ $\rho=-0.66\pm0.01$. We believe that the development of the 27-day GCR wave needs a modulation region of several AUs in the radial direction for all longitudes and that GCR diffusion inside this region leads to the smooth form of the 27-day wave in GCRs.

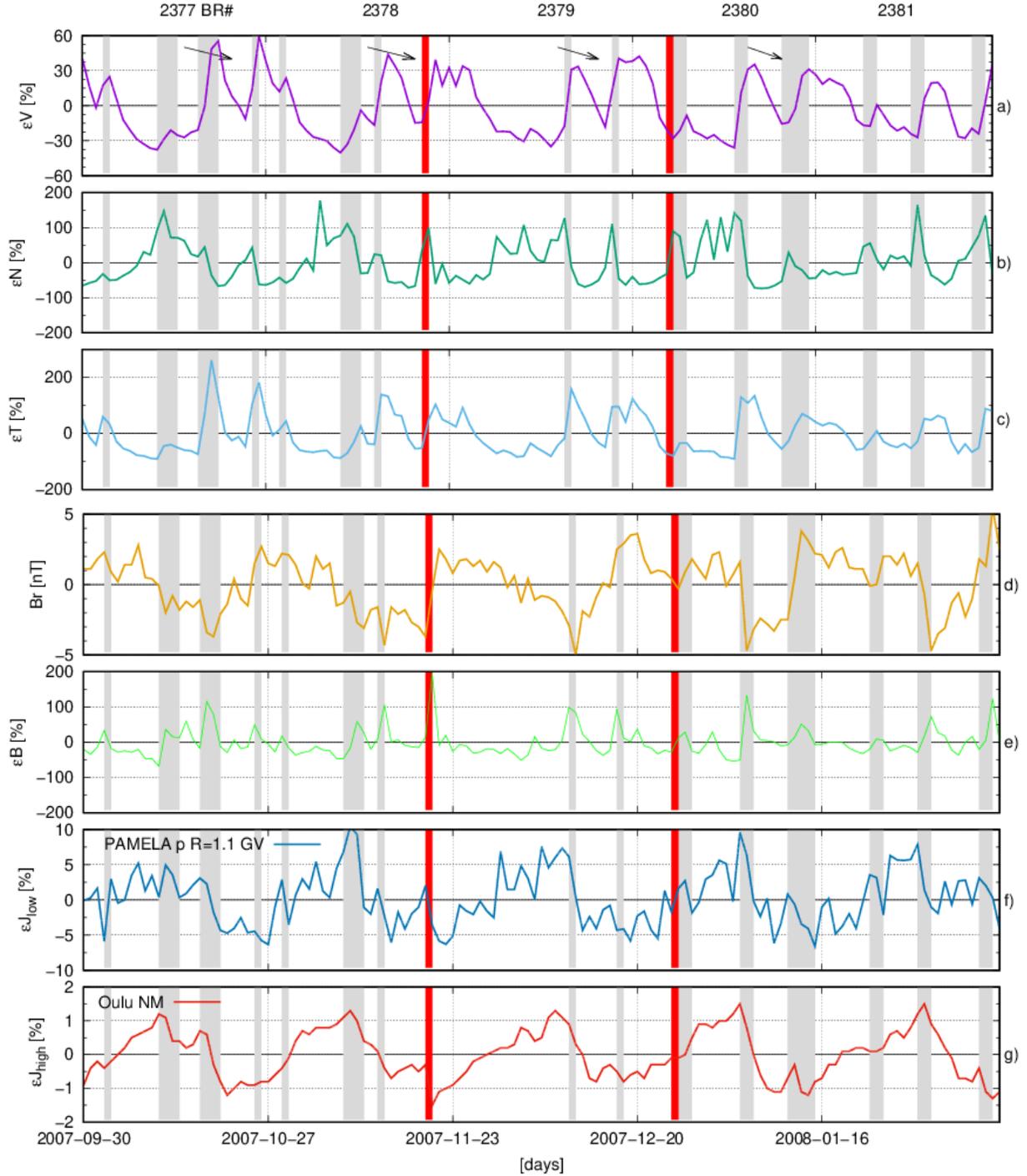

**Figure 5** Relative de-trended 27-day characteristics for Bartels' rotations 2377-2381: (a) SW velocity, **arrows indicate the double SW velocity peaks**, density in (b), temperature in (c), HMF

strength in (e), PAMELA ~1.1 GV protons in (f), and Oulu NM in (g), with panel (d) for the observed HMF *Br* component. SIRs (Jian et al. 2006a, 2011) are shown as vertical shadowed bands and ICMEs (Jian et al. 2006b, 2011) as red bands.

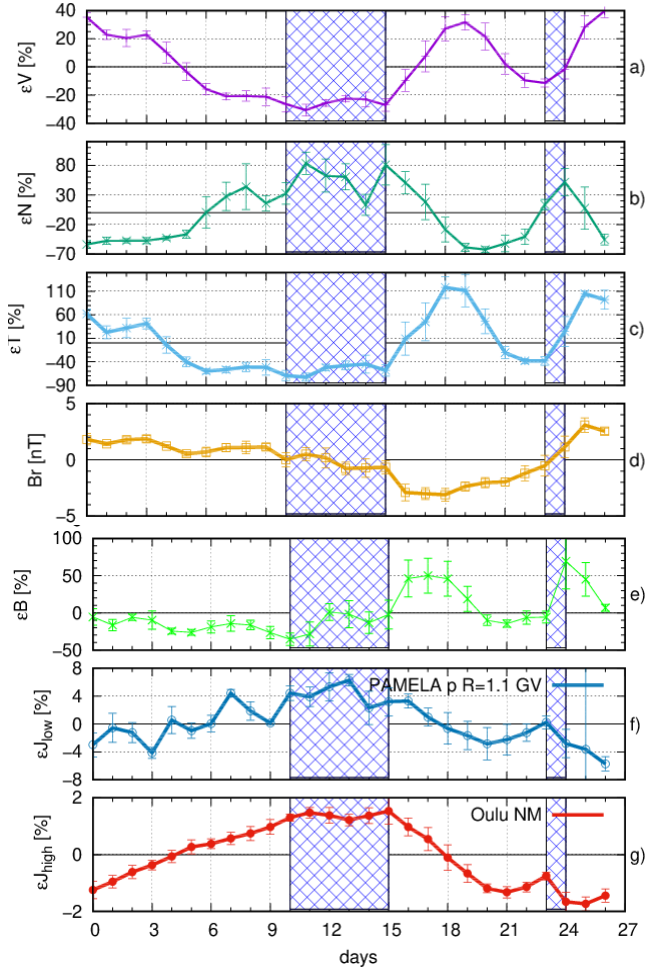

**Figure 6** The 27-day variations in terms of heliospheric parameters (top five panels) and GCRs (as indicated by the legends in two lower panels) averaged over Bartels' rotations 2377-2381. **The passage of the current sheet is shown as vertical marked bands, when *Br* with its errors is approximately zero.**

The relative variation of the value $P_i^j$ of parameter P in the i-th day of the j-th Bartels' rotation with respect of the average value for this BR, $\bar{P}^j$, is calculated as:

$$\epsilon P_i^j = \frac{(P_i^j - \bar{P}^j)}{\bar{P}^j} \cdot 100\%$$

**Then by averaging for five BRs (2377-2381), the mean $\epsilon$Pi and the corresponding error for each parameter is calculated** for each day of the rotation.

Fig. 6 shows the characteristic features of the 27-day variations in terms of GCRs and heliospheric parameters averaged over five Bartels rotations 2377-2381. There are two high-speed streams in the SW velocity with comparable peaks around days 19 and 27 of a Bartels period (panel *a*). There are also two regions of low SW velocity (valleys): a broad one on days ~ 5-17 and a local one on days 21-25. Two enhancements in the HMF strength (panel *e*) formed in the regions where the high-speed streams overtake the low-speed ones. The velocity increase starting on day 23 was accompanied by a short enhancement on day 24 and a subsequent decrease on days 25-27 in the plasma density (panel *b*), an increase of temperature (panel *c*), and HMF strength (panel *e*). All such patterns have the common signature of a SIR. Another velocity increasing, started on day 15, did not bear all features of a SIR as the plasma density displayed a smooth fall and plasma temperature showed a smooth growth. A rather fast HMF sector crossing (*Br*, panel *d*) occurred on day 24, accompanied by a low SW velocity while approaching the sector boundary. As is seen in Fig. 6, panel *d*, the *Br* was rather close to zero from Bartels day 5 to 15. It was marked by the low SW velocity, temperature and enhanced plasma density. This structure (~65% decrease in the velocity from top to bottom) clearly correlates with the enhancement of GCR fluxes (lower panels of Fig. 6, from bottom to top; ~3.5% for the NM and ~12% for *PAMELA* with $R = 1.1$ GV). The comparable velocity decrease from day 19 to 23 (~50%) leads to the small and short corresponding modulation in GCRs (increase on days 22-23; $< \sim 1\%$ for the NM and $< \sim 3.5\%$ for *PAMELA* with $R = 1.1$ GV). Thus, the main puzzle of the relationship between the 27-day GCR variations and the space parameters is the double-peak structure of the SW velocity with the comparable peaks which were not reflected in the GCR fluxes. The latter demonstrates a clear 27-day wave with only one main peak. Perhaps, this issue could be addressed and solved by the realistic modeling of GCR transport under such conditions.

## 5. Rigidity dependence of the 27-day GCR variation

In this section, we study the rigidity dependence of the amplitude of the 27-day GCR variation (A27) for proton fluxes measured by the *PAMELA* and *ARINA* instruments for 2007-2008. Detailed analyses are performed for five BRs 2377-2381 corresponding to 30 September 2007 – 11 February 2008 of the stable 27-day GCR periodicity. The A27 was calculated by means of the normalized and de-trended daily 5-day running average data as the first harmonic of the Fourier extension using harmonic and wavelet methods. For the Fourier method, errors are calculated as the standard deviation from the average value. The amplitudes of the 27-day variations were also derived from a wavelet analysis. The errors of A27 were calculated using the shuffling method (Cassiday et al. 1989, 1990), the relative de-trended data were shuffled 10000 times for each GCR energy range and treated with wavelet procedure. The medians of the simulated A27 distributions were adopted as the A27 errors.

Fig. 7 presents, A27 as a function of the magnetic rigidity *R* for proton fluxes observed by the *PAMELA* and *ARINA* instruments. The figure shows a good agreement between the harmonic and wavelet methods for A27 of the PAMELA proton fluxes.

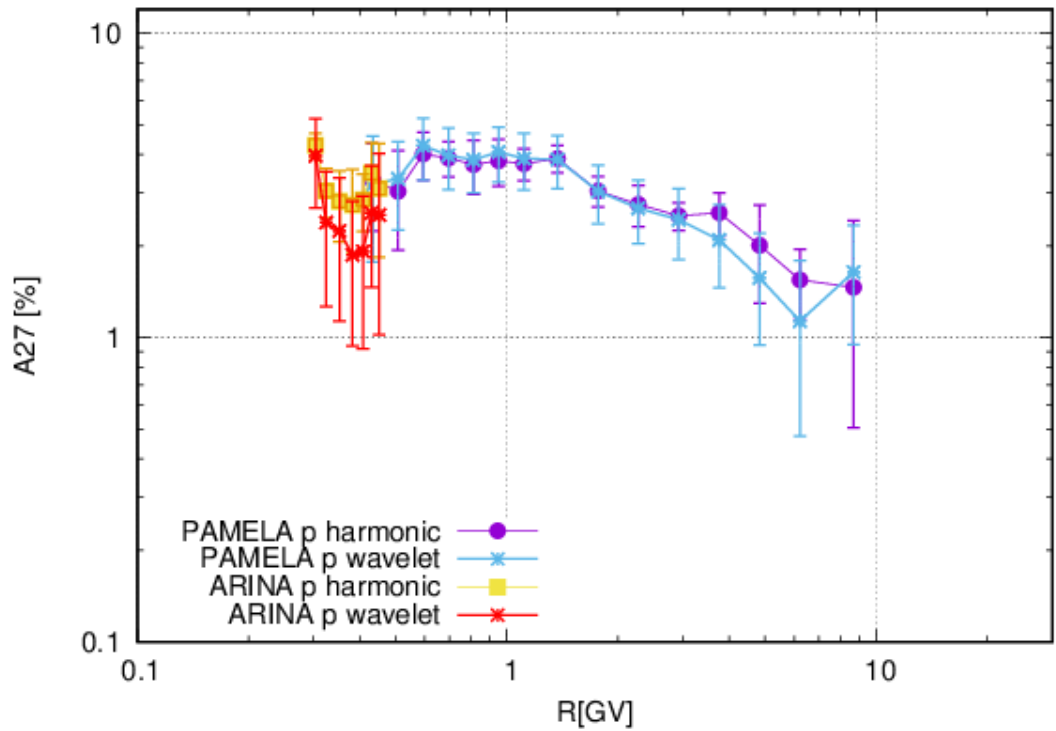

**Figure 7** Rigidity dependence of A27 of proton data observed by the *PAMELA* and *ARINA* instruments; A27 were calculated by harmonic and wavelet methods for 5 BRs 2377-2381 corresponding to 30 September 2007 – 11 February 2008. The solid lines serve as guidance for the eye.

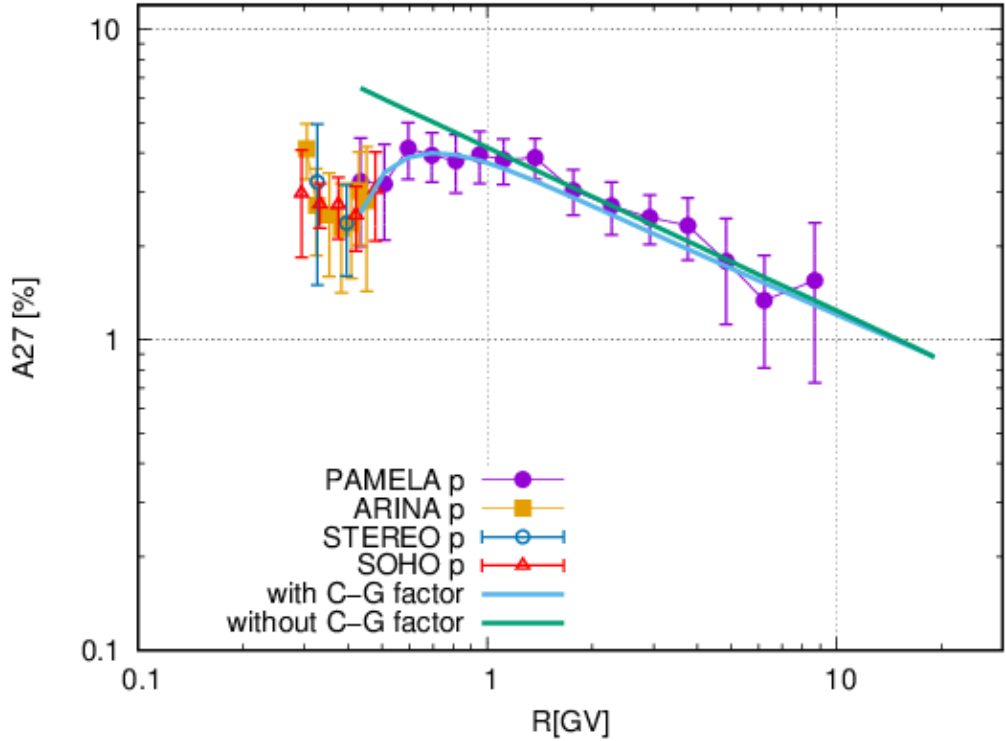

**Figure 8** The A27 for *PAMELA*, *ARINA*, *SOHO* and *STEREO* proton fluxes as a function of rigidity R in 2007-2008. The solid lines through the *PAMELA* and *ARINA* data points are to guide the eye. The spectrum fitting is given when accounting for the Compton-Getting (C-G) effect with the diffusion coefficient $k \sim R^{0.6}$ and without the C-G effect for comparison.

Fig. 7 manifests a non-monotonous form of the A27 rigidity spectrum. For $R \geq 1$ GV it is a power-law with index $\gamma = -0.51 \pm 0.11$ for *PAMELA* protons. However, it flattens significantly when $R < 1$ GV which is described in the interval $R = <0.6 -1>$ GV with a power-law index $\gamma = -0.13 \pm 0.44$. In addition, one can see a local minimum at $R < 0.6$ GV according to these *PAMELA* and *ARINA* results. The growth of A27 at $R < 0.4$ GV which is seen in *SOHO*, *STEREO*, and *ARINA* data is probably caused by the particles accelerated by the CIR-connected shock. Analyzing the rigidity spectrum of the 27-day **GCR variation based on NM observations**, Gil & Alania (2016) found $\gamma = -1.79 \pm 0.09$ in 2007-2008 which does not agree with our finding.

It should be noted that a direct quantitative comparison between the results obtained with NM and *PAMELA* observations is not actually correct. In particular, it is not clear to what energy the NM results relate to because NMs are integral energy detectors, whereas *PAMELA* measurements deal with differential energy bins. Moreover, it should be underlined that although the determination of the effective rigidity of NM was discussed in the literature (e.g. Gil et al. 2017), it is not a simple issue, because the effective rigidity should be different for GCR variations with different rigidity dependence.

To prove that a change in the spectral form indeed occurs, it is important to validate its behavior at $R < 1$ GV against other observations. Leske et al. (2011, 2019) studied the recurrent variations of GCRs and anomalous cosmic rays (ACRs) in the range from 10 to several hundred MeV/n and reported a flatter dependence on energy around 100 MeV/n as in Fig. 5 by Leske et al. (2011) and in Fig. 4 by Leske et al. (2019), which is consistent with our results. Moreover, we have examined the A27 for protons by the *SOHO ERNE HED* experiment [OMNI] (Domingo et al. 1995) for five energy bins from 40-130 MeV/n and *STEREO A* and *B HET* observations [http://www.srl.caltech.edu/STEREO/Public/HET_public.html] for two energy bins from 40-100 MeV/n (Rosenvinge et al. 2008). The results are shown in Fig. 8. Evidently, the presented 27-day amplitudes for *SOHO ERNE* and *STEREO* are consistent with *PAMELA* and *ARINA* measurements for overlapping energy interval.

A clear maximum in A27 around $R \sim 1$ GV was initially observed in 1992-1994 during the out-of-ecliptic journey of the *Ulysses* spacecraft (McKibben et al. 1995). At that time, the spatial and rigidity dependences of the recurrent GCR modulation and the latitudinal GCR gradient showed remarkable similarity (Paizis et al. 1999) and it seemed that there had to be a common modulation process controlling the flux variations in these phenomena. This implies that GCRs with large latitudinal gradients are also strongly modulated by recurrent solar wind structures. Following the approach of Paizis et al. (1999), and to assure compatibility with previously reported experimental work, the force-field approximation of GCR transport (Gleeson & Axford 1968) was implemented as follows:

$$\Delta J/J = -3\, C\, \Delta \Phi,$$

where $\Delta J/J = A27(R)$ and $C$ is the Compton-Getting factor. For the power-law GCR energy spectrum, $dJ/dE \sim E^{-\gamma}$, $C = (2-\alpha\gamma)/3$, $\alpha = (E+2mc^2)/(E+mc^2)$ and $\Phi = \int (V/3k)\, dl$ is the modulation potential, $V$ is solar wind velocity, $k$ is the diffusion coefficient with $dl$ in units of radial distance. Assuming $k \sim R^\delta$, the expression for A27 is

$$\Delta J/J \sim C/(\beta R^\delta),$$

where $\beta$ is the ratio of a particle's speed to the speed of light.

The spectral form of the recurrent GCR variation in 1992-1994 was reasonably well fitted with $\delta = 0.3$- $0.7$ (Paizis et al. 1999). However, the relationship between the 27-day GCR variation and the latitudinal gradient was not corroborated by later observations (Dunzlaff et al. 2008). Moreover, De Simone et al. (2011), based on simultaneous observations by *PAMELA* and *Ulysses* in 2006-2008, found a GCR latitudinal gradient much smaller than predicted in general by first-generation drift models (Jokipii et al. 1977), known to be drift-dominated. However, recent and current drift models reproduce *PAMELA* observations of protons, helium and electrons in detail, as well as the observed radial and latitudinal gradients both in terms of their value (%/AU and %/degree) and their rigidity dependence (Vos & Potgieter 2016). Nevertheless, Leske et al. (2011, 2019), found consistency of the force-field approach (potential modulation model) with the rigidity dependence of the observed recurrent GCR and ACR variation when assuming the diffusion coefficient $k \sim R^{0.5}$.

Following the approach used by Rosenvinge & Paizis (1981), Paizis et al. (1999) and Leske et al. 2011, 2019), we described the rigidity dependence of the 27-day GCR variation retrieved from the *PAMELA* and *ARINA* data, in the framework of the modulation potential concept as described above, which is essentially one-dimensional (1D). So far, sophisticated 3D solar wind and modulation models have not examined the rigidity dependence of 27-day GCR variations (Guo & Florinski 2014, 2016; Wiengarten et al. 2014; Kopp et al. 2017), **except for very recent modeling done by Luo et al. (2020) which is discussed in the next section.**

The key parameter for an explanation of the rigidity dependence of A27 using the modulation potential is a change of $\gamma$ from positive to negative values in the energy range of ~ 150 - 200 MeV/n. The GCR energy spectrum as measured by *PAMELA* (Adriani et al. 2013) was averaged for 6 Carrington rotations 2061-2066 (10 September 2007 – 20 February 2008), which is close to the period of the 27-day episode during the Bartels rotations 2377-2381 (30 September 2007 – 11 February 2008). The GCR energy spectrum did not change essentially during this time period and the power-law index in the energy range of 0.09-14.8 GeV is $\gamma = - 0.771 \ln(E) – 1.1727$. The trends in the observational data on A27 are successfully reproduced provided that the diffusion coefficient $k \sim R^{0.6}$ which is close to the results of Leske et al. (2011). The calculated rigidity spectrum of A27 is shown in Fig. 8 alongside the observational results.

## 6. Discussion

The trans-equatorial coronal holes configuration was stable in 2007-2008 with a ~27-day periodicity (e.g. Gil & Mursula 2018) causing the azimuthally-dependent SW recurrent variations in CIRs. It was shown (e.g. Leske et al. 2011) that in 2007-2008 a remarkable anti-correlation between the SW velocity and GCR occurred. Undoubtedly, the effect of SW streams has an influence on the recurrent GCR modulation discussed above. This effect was not mentioned by Ghanbari et al. (2019), who studied the turbulence properties of the SW around CIRs and suggested that the perpendicular transport is the main source of the GCR modulation around CIRs for both periods, independently of the HMF polarity. Recently, Leske et al. (2019) compared the 27-day variations of GCRs and ACRs in 2007-2008 and 2016-2019 and suggested that a driver of this type variation must involve more than just the latitudinal gradient, explaining it by the fact, that even when the HCS is crossed **more than once** per solar rotation, the clear 27-day variation still persists. Moreover, they found that the direction of the HCS crossings is not a key factor, but more important seems to be whether the observer during the HCS crossings is changing position, that is, leaving the area under the influence of the dominant coronal hole or entering it. These findings show that the competition between convection and the diffusion processes, with changing drift directions in different HMF polarities, should be studied in more detail.

The attempt to connect the 27-day GCR variation with the latitudinal gradient was not successful, but still there are arguments in favor of the influence of the large HMF structure on GCR recurrent variations, that is, the dependence of A27 on the HMF polarity. From a theoretical point of view (e.g. Kota & Jokipii 2001a, 2001b; Burger et al. 2008), GCR A27 is expected to be larger during periods with A > 0 HMF polarity comparing with A < 0 periods. The observational data (e.g. Richardson et al. 1999; Alania et al. 2008) were consistent with this expectation but in 2007-2008 (A<0) A27 (GCR) appears to be nearly equal to those in 1996-1997 (A>0) (Modzelewska & Alania 2012). However, A27(SW) was substantially larger in 2007-2008 than in 1996-1997. Under a reasonable assumption that some relation exists between A27(GCR) and A27(SW), a violation in the relationship between A27(GCR) and HMF polarity might not occur in 2007-2008. Because of the apparent close connection between the GCR recurrent variations and that in the SW velocity, it is reasonable to assume the existence of some proportionality between the A27(GCR) and A27(SW). Considering the periods of the most regular GCR variations as September 1996-April 1997 (Bartels rotations 2227-2235) and October 2007-February 2008 (Bartels rotations 2376-2384) we found that A27(GCR) for the Oulu NM were (0.6 ± 0.07)% and (0.8 ± 0.05)%, respectively. The A27(SW) was substantially less in 1996-1997: (12.2±1.3)% against (20.5±3.3)% in 2007-2008. Therefore, the large A27(GCR) in 2007-2008 could be due to stronger 27-day variations in the SW velocity. Under the equal A27(SW) the recurrent GCR variations in 2007-2008 would be weaker than that in 1996-1997, in agreement with expectations. For example, if A27(SW) in 2007-2008 was (12.2±1.3)% instead of the real (20.5±3.3)%, then A27(GCR) in 2007-2008 would be 0.48% instead of the real 0.8%. It is slightly less than A27(GCR) in 1996-1997 being (0.6±0.07)%.

The relative de-trended GCR intensity in 2007-2008 exhibits a much simpler distribution (one broad minimum and one broad maximum per rotation as shown in Fig. 5) than the SW velocity (two minima of similar depth but of different duration and two close maxima). The smoother distribution of GCRs might be connected with remote GCR modulation by well-developed CIR structures at several AU away. At the same time, the directly observed SW characteristic features at 1AU are local and do not correspond one to one with the observed GCR intensity at Earth. The fully developed and wider CIRs with highly compressed HMF (and hence small diffusion mean free paths) are formed at larger heliocentric distances. Due to the proximity of two CIRs in longitudes with $\partial V/\partial t > 0$ this wide remote diffusion barrier will magnetically correspond to a region encompassing both peaks in $V$ near the Earth. The GCRs traveling to the Earth through this barrier should be more modulated than those traveling through the long valley in the SW velocity. So, we relate the simple distribution of the GCR intensity with the asymmetry of two valleys in SW velocity, corresponding to the asymmetry in the form of the HCS. If this was the case, one important conclusion on the way of interpreting the observed 27-day variation in the GCR intensity is that the real distribution of the SW velocity and HMF in the heliosphere near the Sun should be used as done by Wiengarten et al. (2014), rather than a quite symmetrical case of a simple tilted HCS and the SW velocity depending only on the distance from the HCS as done by Kota & Jokipii (2001b) and Guo & Florinski (2014, 2016).

The rigidity dependence of the GCR recurrent modulation according to the *PAMELA* and *ARINA* data (Figure 7) has a flat maximum in the region of 1 GV and a power-law mode with an index $|\gamma|$ ≈ 0.5 for $R \geq 1$ GV. The latter does not agree with the results of Gil & Alania (2016), for the 2007-2008 episode. The majority of reported research investigated the rigidity dependence of the 27-day GCR variations based on NM observations, which are integral detectors and respond to particle energy above several GeV. We believe that the results by Gil & Alania (2016) may actually relate to $R \geq 15$-20 GV and the spectrum may be steeper in that region. The maximum in the rigidity dependence of the 27-day GCR around 1 GV was found for the first time during the *Ulysses* mission in 1992-1994 (McKibben et al. 1995; Paizis et al. 1999). These results were interpreted in the framework of the force-field modeling approach assuming a rigidity dependence of $k \sim R^\delta$, with $\delta = 0.3$ - 0.7. Applying the same procedure to the P*AMELA* data we get $\delta = 0.6$ in agreement with the previous estimations (Paizis et al. 1999; Leske et al. 2011, 2019). However, the force-field approach does not elucidate the underlying physics.

More promising is a recently reported model by Luo et al. (2020), which used MHD modeling of the SW and HMF as input to a GCR transport code employing a stochastic differential equation approach, where the rigidity dependence of the 27-day GCR variation is also examined. They have conducted a numerical study of the effect of a CIR on the transport of GCR protons. Analyzing the amplitude of GCR variations as introduced by a simulated CIR, it reveals a specific KE dependence of its amplitude variation in the KE range 30 MeV - 2 GeV (rigidity range from 0.24 GV to 2.78 GV). It is noted that in general this simulated KE dependence (Fig. 12 in Luo et al. 2020) is in qualitatively agreement with *PAMELA* observations as demonstrated in our study, that is, the amplitude-rigidity-energy profile exhibits a flat interval in the lower range (KE < 200 MeV), but then decreases as KE increases above this range. According to Luo et al. (2020), the magnetic field *B* is enhanced inside the CIR, so if the diffusion coefficients scale spatially as $\frac{1}{B}$ ($k \sim \frac{1}{B}$), the CIR essentially acts as a diffusion barrier for GCRs in their numerical transport model. This means that at larger radial distances the CIR effect on GCRs will become smaller. According to previously done 1D diffusion barrier modeling (le Roux & Potgieter 1989), the value of the GCR intensity decrease is given by $\frac{\Delta J}{J} \sim \frac{\Delta k}{k^2}$; if this $k$ increases with increasing KE (rigidity), the decrease amount will become progressively smaller with increasing KE. In 3D modeling, this $k$ actually consists of three dissimilar diffusion coefficients, as well as a drift coefficient, which all may have a different rigidity dependence (see Fig. 10 in the review by Potgieter 2017), apart from having a latitudinal and longitudinal dependence. This means that the amount of decrease in GCR intensity may change in a more complicated quantitative manner with increasing KE (rigidity). Nevertheless, conceptually and qualitatively, the features of GCR variations caused by CIRs, should not change significantly in 3D modeling. Evidently, at this stage the 3D numerical model presented by Luo et al. (2020) cannot fully reproduce the A27(GCR) rigidity dependence following from the *PAMELA* observations reported here, which reveal a much broader flatness for its rigidity dependence. Furthermore, the exact physical processes causing this "flat and steady" rigidity dependence still seem somewhat ambiguous. However, several factors may contribute to the quantitative discrepancy. There may be technical differences, for example, the simulation from Luo et al. (2020) is not presented at the Earth but at 3.01 AU; and perhaps also the manner in which they defined and calculate the 'amplitude' of the GCR variation. It is noted that the three diffusion coefficients in their numerical model are of a simplified form, essentially all of them scale as $R^{0.5}$. In this context, if their diffusion coefficients had a less strong dependence on rigidity at lower rigidities, the compatibility between the simulations and observations could be better. Additionally, the physical extent (dimensions) of the simulated CIR also plays an important role, in terms of a radial, latitudinal and longitudinal (azimuthal) dependence, and how this may evolve with time. For this type of modeling, in order to reproduce in detail the observed A27(GCR) rigidity dependence in the inner heliosphere, it seems required that the parameter space for GCR modulation should be explored in detail, especially the functional dependence of all relevant diffusion coefficients on rigidity. In this way, we may contribute to a better understanding of physical processes governing the GCR 27-day variation.

## 7. Conclusions

1. This study reports for the first time the rigidity dependence of the amplitude of the 27-day variation (A27) of the GCR intensity observed directly in space in a wide rigidity range from ~0.3 GV to ~10 GV as observed by the space-borne instruments *PAMELA* and *ARINA*. The rigidity dependence of A27 (GCR) cannot be described by the same power-law at both low and high rigidities. The rigidity spectrum of A27 (GCR) manifests a non-monotonous form; for $R \geq 1$ GV a power-law dependence is noticeable with index $\gamma = -0.51 \pm 0.11$ for *PAMELA* protons; for $R = \langle 0.6-1 \rangle$ GV the rigidity dependence becomes flatter with $\gamma = -0.13 \pm 0.44$ for *PAMELA* protons. According to *PAMELA* and *ARINA* results, a local minimum in the rigidity dependence for $R < 0.6$ GV is present. Such a flatter dependence of A27 (GCR) on energy around 100 MeV/n is also found for *SOHO ERNE* and *STEREO*, being consistent with *PAMELA* and *ARINA* measurements for overlapping rigidity intervals.

2. The Compton-Getting effect was previously applied by Paizis et al. (1997) for the explanation of the rigidity dependence of recurrent variations in *Ulysses* data and recently by Leske et al. (2011; 2019) for *ACE* measurements. We also have used this approach effectively when analyzing *PAMELA* data and found the force-field diffusion coefficient $k \sim R^{0.6}$.

3. The time lines of the SW velocity, the HMF and GCR intensity near the Earth in the selected period of intense 27-day variations in 2007-2008 are quasi-steady but longitudinally dependent in the coordinate system rotating with the Sun. A remarkable anti-correlation was observed between the SW velocity and GCRs. However, the GCR intensity for both low and high rigidities demonstrates a smoother 27-day wave in comparison with the SW velocity recurrence. It clearly shows that the GCR variation does not reflect the structural features of the local heliospheric conditions observed near the Earth. We believe that the development of the 27-day GCR wave requires a modulation region of several AU in the radial direction for all longitudes, and that GCR diffusion inside this region leads to the smooth form of the 27-day wave in GCR.

4. Results of MHD modelling of the SW plasma and stochastic transport model of GCR protons around CIRs by Luo et al. (2020) are qualitatively in agreement with *PAMELA* observations i.e., the rigidity dependence of the amplitude exhibits a flat interval for KE < 200 MeV but then decreases as KE increases above this range. It is caused by the enhanced magnetic field *B* inside the CIR, so if the diffusion coefficients scale spatially as $\frac{1}{B}$ ($k \sim \frac{1}{B}$), the CIR essentially acts as a diffusion barrier for GCRs. Nevertheless, in order to reproduce quantitatively the observed rigidity dependence of the 27-day GCR amplitude, a more sophisticated study of GCR modulation in and around CIRs is required, especially the functional dependence of the diffusion coefficients on rigidity. **Such a model can also be used to investigate if drifts play a meaningful role in these observations. In this context, the rigidity dependence of the 27-day GCR variation that follows from observations could be helpful.** Moreover, the flattening of the rigidity dependence of the observed 27-day amplitude, may indicate that the rigidity dependence of the diffusion coefficients is also flattening off with decreasing rigidity and the rigidity range where this happens can be determined experimentally.

**Appendix**

The PAMELA proton fluxes were evaluated on a daily basis between January 2007 and December 2008. The proton sample was obtained with the selection cuts already described in (Adriani et al. 2013; Martucci et al. 2018), hence no isotopic separation (proton/deuterium) was performed. The efficiencies were also evaluated following the same approach.

Time-dependent flux measurements require precise evaluation of any efficiency variation over time. Between 2007 and 2008, only the tracker efficiency was found to have a significant time dependence, therefore a dedicated analysis was developed for its determination. The tracker selection efficiency was evaluated combining the simulated and flight information as previously done in Adriani et al. (2011) and in Adriani at al. (2013). First, the Monte Carlo simulation was used to estimate the tracker efficiency and its dependences over rigidity and time. The simulation of the PAMELA tracking system accounted for the measured noise of each silicon plane and for the performance variations over the duration of the measurement also by including a map of dead channels over time. The Monte Carlo approach was also validated estimating the tracker selection efficiency using simulated and experimental events classified as non-interacting minimum-ionizing particles by the imaging calorimeter (i.e. mostly protons with rigidities greater than 2 GV). A good agreement between the resulting simulated and experimental tracker selection efficiencies was found.

Lastly, in order to evaluate any residual time dependence, the high-energy daily fluxes obtained in this analysis were compared to the high-energy flux measured over the period July 2006 - March 2008 (Adriani et al. 2011). Residual systematic time-dependence effects were found and were corrected with normalization factors obtained normalizing the daily fluxes between 20 GeV and 50 GeV to the Adriani et al. (2011) flux (lowered by 3.2% as indicated in Adriani et al. 2013) measured in the same rigidity region. To reduce the statistical fluctuations of these daily normalization factors, a running average value, with a window of 20 days, was calculated and used to normalize the daily fluxes. The resulting daily proton rigidity spectra measured by PAMELA from January 2007 until December 2008 are shown in Figure 1A. **The PAMELA data discussed in this paper will be available at the Cosmic Ray Data Base of the ASI Space Science Data Center (http://tools.asdc.asi.it/CosmicRays/chargedCosmicRays.jsp).**

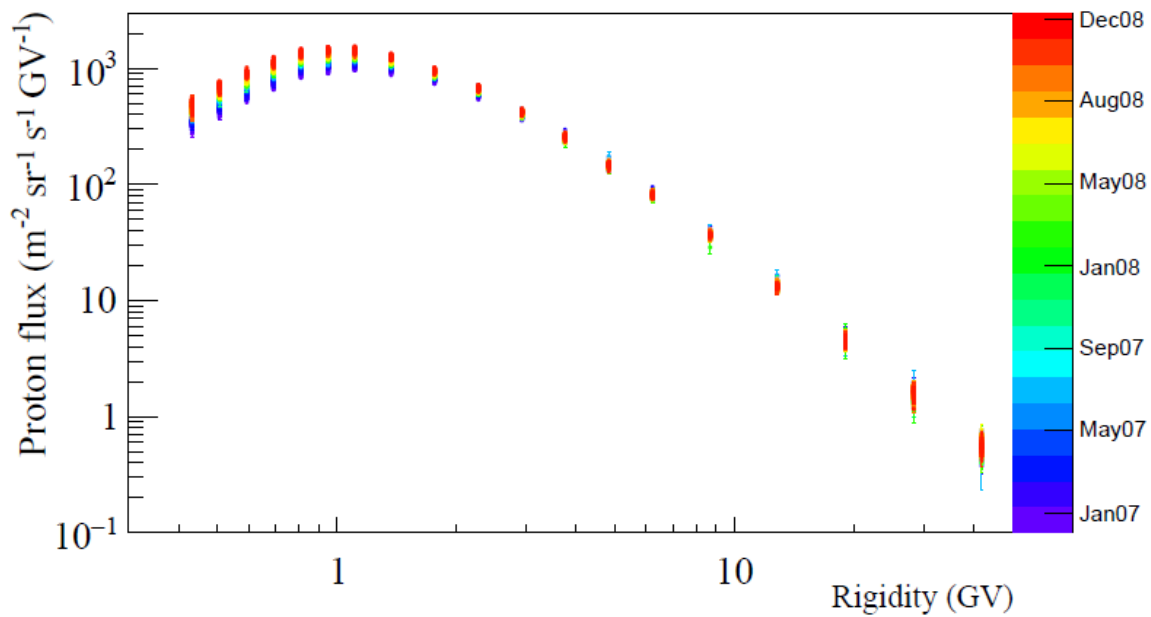

**Figure 1A** Daily proton rigidity spectra measured by PAMELA from January 2007 until December 2008

**Acknowledgments**

We acknowledge partial financial support from The Italian Space Agency (ASI) under the program "Programma PAMELA - attivita' scientifica di analisi dati in fase E". We also acknowledge support from Deutsches fur Luft- und Raumfahrt (DLR), The Swedish National Space Board, The Swedish Research Council, The Russian Space Agency (Roscosmos), Russian Science Foundation and NASA Supporting Research Grant 13SRHSPH1320075, the RFBR grant 18-02-00582a, RFBR and NRF grant 19-52-60003 SA-t. Oulu Neutron monitor count rates are from https://cosmicrays.oulu.fi/. Data of heliospheric parameters are from OMNI

[http://omniweb.gsfc.nasa.gov]. We acknowledge *SOHO ERNE HED* [data from OMNI] and *STEREO A, B HET* [http://www.srl.caltech.edu/STEREO/Public/HET_public.html] experiments. R.M. acknowledges The Polish National Science Centre, decision number 2017/01/X/ST9/01023. R.M. acknowledges partial financial support from the INFN Grant "giovani", project ASMDM. X.L. was partially supported by the NSFC grants (41774185, U1738128) and the Shandong Institute of Advanced Technology. The work has been performed with a support of the Ministry of Science and Higher Education of the Russian Federation, Project "Fundamental problems of cosmic rays and dark matter", No 0723-2020-0040.